\documentclass[12pt]{iopart}

\usepackage{cite}

\begin{document}

\title[Kratzer potential in hyperspherical coordinates]
{Comment on: ``Non--relativistic treatment of diatomic molecules
interacting with generalized Kratzer potential in hyperspherical
coordinates''}

\author{Francisco M Fern\'andez}

\address{INIFTA (UNLP, CCT La Plata--CONICET), Divisi\'on Qu\'imica Te\'orica,
Blvd. 113 S/N,  Sucursal 4, Casilla de Correo 16, 1900 La Plata,
Argentina}\ead{fernande@quimica.unlp.edu.ar}

\maketitle

\begin{abstract}
We argue that the textbook method for solving eigenvalue equations
is simpler, more elegant and efficient than the Asymptotic
Iteration Method (AIM) applied in J. Phys. A {\bf 44} 155205. We
show that the Kratzer potential is not a realistic model for the
vibration--rotation spectrua of diatomic molecules because it
predicts the position of the absorption infrared bands too far
from the experimental ones (at least for the $HCl$ and $H_2$
molecules chosen as illustrative examples in that paper).
\end{abstract}

In order to study the vibration--rotation motion of diatomic
molecules in $N$ dimensions Durmus\cite{D11} chose the Kratzer
potential and solved the Schr\"{o}dinger equation in
hyperspherical coordinates by means of the Asymptotic Iteration
Method (AIM). He obtained the well known results and as an
illustrative and practical application of the model he restricted
himself to the only apparently relevant case $N=3$ calculating
some vibration--rotation energies for $H_{2}$ and $HCl$. In what
follows we contrast the AIM derivation of the main equations with
the well known and widely used textbook approach, and also compare
the theoretical results for those molecules with experimental
ones.

The starting point is the Schr\"{o}dinger equation
\begin{equation}
\left[ -\frac{\hbar ^{2}}{2\mu }\nabla ^{2}+V(r)\right] \Psi (\mathbf{r}%
)=E\Psi (\mathbf{r})  \label{eq:Schro}
\end{equation}
where $\mu $ is the reduced mass of the molecule and $E$ is the
vibration--rotation energy. As a ``realistic'' model for the
interaction between the nuclei the author chose the Kratzer
potential
\begin{equation}
V(r)=D_{e}\left( \frac{r-r_{e}}{r}\right) ^{2}+\eta   \label{eq:V_Kratzer}
\end{equation}
where $r_{e}$ is the equilibrium internuclear separation and
$D_{e}$ is the molecular dissociation energy (misleadingly called
intermolecular separation and dissociation energy between diatomic
molecules, respectively, by the author\cite{D11}). He made a
curious distinction between the modified Kratzer potential $\eta
=0$ and the Kratzer potential $\eta =-D_{e}$ which are just two
alternative expressions of the same interaction with the energy
origin shifted by $\eta $. Without a plausible justification he
further argued that it is useful from a physical point of view to
consider the general $N$--dimensional case where
\begin{equation}
r^{2}=\sum_{j=1}^{N}x_{j}^{2},\;\nabla ^{2}=\sum_{j=1}^{N}\frac{\partial ^{2}%
}{\partial x_{j}^{2}}  \label{eq:N_dim}
\end{equation}

Since the potential is spherically symmetric one can separate the Schr\"{o}%
dinger equation (\ref{eq:Schro}) into its radial and angular parts, and by
means of well--known transformations the former is reduced to\cite{D11}
\begin{equation}
yF^{\prime \prime }(y)+\left( 2\gamma +N-1-2\beta y\right) F^{\prime
}(y)+\left[ 2\kappa -\beta (2\gamma +N-1)\right] F(y)=0  \label{eq:F(y)}
\end{equation}
where
\begin{eqnarray}
y &=&\frac{r}{r_{e}}  \nonumber \\
\kappa  &=&\frac{2\mu D_{e}r_{e}^{2}}{\hbar ^{2}}  \nonumber \\
\beta ^{2} &=&\frac{2\mu r_{e}^{2}(D_{e}-E+\eta )}{\hbar ^{2}}  \nonumber \\
\gamma  &=&\frac{N-2}{2}+\sqrt{\kappa +\left( l+\frac{N-2}{2}\right) ^{2}}
\label{eq:parameters}
\end{eqnarray}
and $l=0,1,\ldots $ is the angular--momentum quantum number
(called $l_{N-1}$ in Ref.~\cite{D11}). In order to solve this
equation Durmus\cite{D11} applied the AIM which is an iterative
approach that gives results for $n=0,1,\ldots $, where $n$ labels
the number of iterations. By inspection of the particular outputs
$\beta _{0l},\,\beta _{1l},\ldots $ one hopefully derives the
value of $\beta _{nl}$ for an arbitrary number $n$ of iteration
steps and then the allowed energy $E=E_{nl} $. With some more
ingenuity one realizes that $F(y)$ is proportional to the
confluent hypergeometric function
${\vphantom{F}}_{1}{F}_{1}(-n,2\gamma +N-1;2\beta y)$. This
procedure offers little difficulty if one already knows the exact
result beforehand which is actually the case here. Once we have
the solution in terms of the confluent hypergeometric function we
easily rewrite it in terms of the associated Laguerre
polynomials\cite{D11,AS72}.

Solving equation (\ref{eq:F(y)}) is a textbook problem\cite{F99}
and the widely known approach is faster, more elegant and
efficient than the AIM. If we define $y=\frac{z}{2\beta }$ then
$w(z)=F(\frac{z}{2\beta })$ satisfies
\begin{equation}
zw^{\prime \prime }(z)+\left( 2\gamma +N-1-z\right) w^{\prime }(z)+\left[
\frac{\kappa }{\beta }-\left( \gamma +\frac{N-1}{2}\right) \right] w(z)=0
\label{eq:w(z)}
\end{equation}
that is a particular case of Kummer's equation\cite{AS72}
\begin{equation}
zw^{\prime \prime }(z)+\left( b-z\right) w^{\prime }(z)-aw(z)=0
\label{eq:Kummer_eq}
\end{equation}
The Kummer's function can be expanded in a power--series
\begin{equation}
M(a,b,z)=\sum_{j=0}^{\infty }\frac{a_{j}z^{j}}{j!b_{j}}  \label{eq:Kummer_M}
\end{equation}
where $\xi _{j}=\xi (\xi +1)(\xi +2)...(\xi +j-1)$ and $\xi _{0}=1$. It
becomes a polynomial of degree $n$ when $a=-n$ and $b\neq -m$ ($m$ and $n$
positive integers). On comparing equations (\ref{eq:w(z)}) and (\ref
{eq:Kummer_eq}) we directly obtain
\begin{equation}
\beta _{nl}=\frac{\kappa }{n+\gamma +\frac{N-1}{2}}  \label{eq:beta_n}
\end{equation}
and\cite{AS72}
\begin{equation}
F_{n}(y)=M(-n,2\gamma +N-1,2\beta y)={\vphantom{F}}_{1}{F}_{1}(-n,2\gamma
+N-1;2\beta y)
\end{equation}
It is clear that one can derive the solution to equation
(\ref{eq:F(y)}) directly from comparison of the appropriate
equations in a way that makes the AIM utterly unnecessary.

Durmus\cite{D11} did not show any physical application of the
$N$--dimensional model except for the obvious case $N=3$. Most
curiously he showed results in the form of a table and figure for
both $\eta =0$ and $\eta =-D_{e}$. Apparently, he did not realize
that both spectra are the same but for an energy shift that is
irrelevant from a physical point of view. Suffice to say that the
physical observables are not affected by this shift.

Another curious fact is that in Table 2 the author only showed the
energies for $n\geq $ $l=0,1,\ldots $. In the case of a diatomic
molecule, to which the model is supposed to apply, $n$ is the
vibrational quantum number $v$ and $l$ is the rotational quantum
number $J$ so that such selection of quantum numbers is of scarce
utility from a spectroscopic point of view (one would expect
increasing $J$ for a given $v$).

Durmus\cite{D11} argued that the Kratzer potential provides a realistic
description of molecular vibrations, but it is far from true as discussed by
Pl\'{i}va\cite{P99} who stated that ``However, in its basic form, this
function provides only a rather crude approximation for the molecular
potential, and for this reason it has not been popular with
spectroscopists''. In the first place, the Kratzer potential supports an
infinite number of vibration--rotation levels which is not the case of
actual diatomic molecules. In addition to it, it does not describe the
spectrum correctly as we will show in what follows. From the
vibration--rotation energies written in the usual spectroscopic way
\begin{equation}
E_{v,J}=-\frac{\kappa D_{e}}{\left[ v+\frac{1}{2}+\sqrt{\kappa
+\left( J+\frac{1}{2}\right) ^{2}}\right] ^{2}}  \label{eq:EvJ}
\end{equation}
we obtain the spectral lines in wavenumber units
\begin{equation}
\tilde{\nu}=\frac{E_{\nu ^{\prime },J^{\prime }}-E_{v,J}}{hc}  \label{eq:nu}
\end{equation}
according to the selection rules $\Delta v=\pm 1$ and $\Delta
J=\pm 1$.\cite {H50} The results in Durmus' Table 2 do not even
allow us to obtain the P and R branches of the fundamental
absorption band ($v=0$, $v^{\prime }=1$, $J^{\prime }=J\pm
1$).\cite{H50} The first two entries for $HCl$ in that table
predict the center of this band to appear at
$E_{1,0}-E_{0,0}=0.1482\,eV$ or $1195\,cm^{-1}$. However, it is
well known that the center of the fundamental band is located at
$\tilde{\nu}=2886\,cm^{-1}$,\cite{H50} more than twice the value
given by Durmus' energies. It is not better for $H_2$
(theoretical$=2715.7\,cm^{-1}$, experimental$=4160.2\,cm^{-1}$) as
expected from a model that no spectroscopist would take seriously.

Summarizing the main conclusions of this comment we may say that
the application of the AIM for solving the Schr\"{o}dinger
equation with the Kratzer potential is far from being the best
strategy. One obtains the same expressions in a more direct, easy
and elegant way by means of the standard textbook
method.\cite{F99} In addition to it, since Durmus did not show any
plausible physical application of the model to space dimensions
other than $N=3$ and the results for ordinary diatomic molecules
are extremely poor we conclude that the model is of scarce
physical utility.

\end{document}